\documentclass[]{mn2e}
\usepackage{epsfig}
\usepackage{times}
\def\spose#1{\hbox to 0pt{#1\hss}}
\newcommand\lsim{\mathrel{\spose{\lower 3.0pt\hbox{$\mathchar"218$}}
     \raise 2.0pt\hbox{$\mathchar"13C$}}}
\newcommand\gsim{\mathrel{\spose{\lower 3.0pt\hbox{$\mathchar"218$}}
     \raise 2.0pt\hbox{$\mathchar"13E$}}}
\newcommand\msun{{\rm \,M_\odot}}

\title[Evolution of the Globular Cluster System]{Dynamical Evolution of
the Mass Function and Radial Profile of the Galactic Globular Cluster System}
\author[J. Shin, S. S. Kim, and K. Takahashi]{Jihye Shin$^{1}$, Sungsoo S.
Kim$^{1}$\thanks{Corresponding author: sungsoo.kim@khu.ac.kr},
and Koji Takahashi$^{2}$\\
$^{1}$Department of Astronomy and Space Science, Kyung Hee
University, Kyungki 446-701, Korea; jhshin@ap4.khu.ac.kr\\
$^{2}$Department of Informational Society Studies, Saitama
Institude of Technology, Fukaya, Saitama 369-0293, Japan}

\begin{document}

%\date{Accepted. Received; in original form}
%\pagerange{\pageref{firstpage}--\pageref{lastpage}} \pubyear{2002}

\maketitle
\label{firstpage}

\begin{abstract}
Evolution of the mass function (MF) and radial distribution (RD) of the
Galactic globular cluster (GC) system is calculated using an advanced
and realistic Fokker-Planck (FP) model that considers dynamical friction,
disc/bulge shocks, and eccentric cluster orbits.  We perform hundreds of
FP calculations with different initial cluster conditions, and then search
a wide parameter space for the best-fitting initial GC MF and RD that evolves
into the observed present-day Galactic GC MF and RD.  By allowing both
MF and RD of the initial GC system to vary, which is attempted for
the first time in the present {\it Letter}, we find that our best-fitting
models have a higher peak mass for a lognormal initial MF and a higher
cut-off mass for a power-law initial MF than previous estimates, but
our initial total masses in GCs, $M_{T,i}=1.5$--$1.8 \times 10^8 \msun$, are
comparable to previous results.
Significant findings include that our best-fitting lognormal MF shifts
downward by 0.35 dex during the period of 13 Gyr, and that our power-law
initial MF models well-fit the observed MF and RD only when the initial MF
is truncated at $\gsim 10^5 \msun$.
We also find that our results are insensitive to the initial distribution
of orbit eccentricity and inclination, but are rather sensitive to the
initial concentration of the clusters and to how the initial tidal radius
is defined.
If the clusters are assumed to be formed at the apocentre while filling 
the tidal radius there, $M_{T,i}$ can be as high as $6.9 \times 10^8 \msun$,
which amounts to $\sim 75$ per cent of the current mass in the stellar halo.
\end{abstract}


%%%%%%%%%%%%%%%%%%%%%%%%%%%%%%%%%%%%%%%%%%%%%%%%%%%%%%%%%%%%%%%%%%%%%%%%%%%%%%%%
\section{INTRODUCTION}
\label{sec:intro}

Globular clusters (GCs) are the oldest ($\sim 13$ Gyr) bound stellar
subsystems in the Milky Way, and studies on the evolution of the GCs
may give us valuable information on the environment of the Milky Way
at the era of its formation.  The initial mass function (MF) of the GC
system is particularly of interest since it can tell us about the mode of
star formation and about the fractions of stars that are formed inside
and outside the clusters at the beginning of the galaxy.
%The present-day globular cluster mass function (GCMF) of the Milky Way
%is indeed quite different from those of young cluster systems in other
%galaxies.  The Galactic GC system presently has a lognormal MF
%with a peak at $M_p= 2 \times 10^5 M_\odot$ and a dispersion of
%$\sigma (\log M/ M_\odot)=0.5$, while, for example, the MF of young cluster
%system in merging Antennae galaxies appears to follow a simple power-law
%function (Zhang \& Fall 1999).
%These facts led to a suggestion that the Galactic GCMF
%was initially a power-law and has evolved into a lognormal due to
%the dynamical evolution and disruption of individual GCs (e.g., Fall \& Rees
%1977).  Later, lognormal initial GCMFs (hereafter IGCMFs) began to be
%considered as well (e.g., Fall \& Rees 1985), and recently,
%Parmentier \& Gilmore (2007) find that a power-law initial mass distribution
%of protoglobular clouds can quickly evolve into a lognormal initial GCMF due
%to the expulsion of the gas remnant from star formation.

Evolution of the globular cluster mass function (GCMF\footnote{By the MF
of the GC system or GCMF, we mean the number of clusters, not stars, as a
function of mass.}) is driven by many
factors such as two-body relaxation, stellar evolution, binary heating,
galactic tidal field, eccentric cluster orbits, and disc/bulge shocks.
It is quite challenging to calculate
the evolution of the GCMF considering all these factors.  There have been
numerous studies on the evolution of the GCMF using analytical models
(Aguilar, Hut, \& Ostriker 1988; Okazaki \& Tosa 1995; Vesperini 1997;
Fall \& Zhang 2001, among others), Fokker-Planck (FP) models (Gnedin \&
Ostriker 1997; Murali \& Weinberg 1997, among others), and N-body models
(Vesperini \& Heggie 1997; Baumgardt 1998; Vesperini 1998, among others),
but none of these studies took all of the
aforementioned physics into account or implemented a wide enough range of
parameter space for the initial conditions of the GCs.  It is rather
difficult to incorporate all these physics into analytical or FP models,
whereas N-body simulations, although generally more accurate and easier to
consider all the disruption mechanisms than the former, are still too
CPU-expensive to be performed for clusters with $N \gsim 10^5$.

In the present {\it letter}, the evolution of the Galactic GCMFs is calculated
using the most advanced, realistic FP model developed so far that
incorporates all of the disruption mechanisms discussed earlier.  We perform
FP calculations for 720 different initial conditions (mass, galactocentric
radius, orbit eccentricity, and orbit inclination).  We then search
a wide-parameter space for the best-fitting initial GC mass and radial
distributions (RDs) that evolve into the observed present-day Galactic GC
distributions (MF and RD).  Such a simultaneous fit to both MF and RD
of the GC system is attempted for the first time in the present study.
We adopt a lognormal and a truncated power-law function for the
initial GCMF (IGCMF)\footnote{Parmentier \& Gilmore (2007) find
that a power-law initial
mass distribution of protoglobular clouds can quickly evolve into a
lognormal initial GCMF due to the expulsion of the gas remnant from
star formation, if the power-law mass distribution has a lower mass limit}.
Our IGCMFs are to be regarded as the models after
the gas expulsion., and a softened power-law function for the initial
RD of the apocentre.

%%%%%%%%%%%%%%%%%%%%%%%%%%%%%%%%%%%%%%%%%%%%%%%%%%%%%%%%%%%%%%%%%%%%%%%%%%%%%%%%
\section{THE PRESENT-DAY GCMF} 

The Galactic GCs can be classified into three groups by their age and
metallicity: the `old' halo (OH) and bulge/disc (BD) clusters are believed to
be Galactic natives that were created when the protogalaxy collapsed, 
while the `young' halo (YH) clusters are thought to have been formed in
external satellite galaxies (Zinn 1993; Parmentier et al. 2000).
When comparing our calculations to the present-day GCMF, we only consider
the `native' clusters, i.e., the OH and BD clusters.
We adopt the GC classification by Mackey \& van den Bergh (2005),
which are based on the data base compiled by Harris (1996).
Our native GCs do not include the six objects that belong to the
Sagittarius dwarf, seven objects whose origins remain unknown
(Mackey \& van den Bergh 2005), and 15 objects that are thought to be
the remnants of dwarf galaxies (Lee, Gim, \& Casetti-Dinescu 2007).
The total number of the native clusters is 95 (61 OHs and 34 BDs), and
the total mass in these clusters is $2.6 \times 10^7 \msun$
($1.8 \times 10^7 \msun$ in OHs and $0.8 \times 10^7 \msun$ in BDs).
When fit to a lognormal function, the MF of our native clusters has
$M_p=10^{5.26} \msun$ and $\sigma_{\log M}=0.44$.
The observed luminosities of the GCs are transformed into masses with
a mass-to-light ratio of $M/L_V=2$.

%%%%%%%%%%%%%%%%%%%%%%%%%%%%%%%%%%%%%%%%%%%%%%%%%%%%%%%%%%%%%%%%%%%%%%%%%%%%%%%%
\section{MODELS AND INITIAL CONDITIONS}
\label{sec:model}

We adopt the anisotropic FP model developed by Takahashi \& Lee (2000)
and Takahashi \& Portegies Zwart (2000),
which directly integrates the orbit-averaged FP equation of two
(energy-angular momentum) dimensions and considers multiple stellar
mass components and the effects of tidal fields, three-body binary heating
and stellar evolution.  To this model, we have added the effects of
tidal binary heating, disc/bulge shocks, dynamical friction and realistic
cluster orbits.

Disk and bulge shocks arise when clusters pass through the galactic disc
or the bulge.  Shocks inject kinetic energy into the cluster and
speedup its disruption.  Gnedin, Lee, \& Ostriker (1999) incorporated
these shocks into an FP model of one (energy) dimension.  We have extended
their recipe and applied it to our two-dimensional (2D) FP model (detailed
description on this application will be presented in Shin, Kim, \& Takahashi
2008).  The FP model is numerically stable in most cases, but we find that
it encounters numerical problems rather often when the effects of tidal
shocks are included in the anisotropic FP model.  To avoid such a problem,
Shin \& Kim (2007) developed a new integration scheme for a 2D
FP equation by adopting an Alternating Direction Implicit method.  We
use this scheme for our calculations.

Dynamical friction between a cluster and galactic field stars gradually
transports a cluster to the inner region of the galaxy.  Since a cluster
with a given mass $M$ has a smaller tidal radius when located at a smaller
galactocentric radius $R$, dynamical friction increases the mass loss rate
$\dot M$ of the cluster over its tidal radius $r_t$.  Eccentric cluster
orbits have similar effects on the clusters, although the effects are
transient and periodic.  To incorporate the
effects of dynamical friction and eccentric orbits into our FP model,
we follow the orbit of the cluster by integrating the equation of motion
in the Galactic potential with a drag due to dynamical friction, and
continuously update $r_t$ of the cluster at each time-step, which is
determined by the current $R$ and $M$.
\footnote{It appears that such a
continous update of $r_t$ with a realistic orbit calculation is
the first ever attempt for FP models.}
For a drag due to dynamical friction, we adopt the formula by Chandrasekhar
(1943), and for the Galactic potential, we employ the model by Johnston,
Spergel, \& Hernquist (1995).

Takahashi \& Portegies Zwart (1998, 2000) found that FP models produce
results similar to those from N-body simulations at least for clusters
on circular orbits, if an ``apocenter criterion'' and $\nu_{esc}=2$--3
are used for the escape criterion of the FP model ($\nu_{esc}$ is a
dimensionless parameter that determines the time-scale on which escaping
stars leave the cluster).  We adopt the apocenter criterion as well and
$\nu_{esc}=2.5$.
We find that the FP and N-body methods show a good agreement for clusters
on eccentric orbits as well.  A comparison between our FP calculations and
N-body simulations of Baumgardt \& Makino (2003) for clusters
on eccentric orbits with initial masses larger than $10^4 \msun$ shows
a good agreement of cluster lifetimes within $\sim 25$ per cent.

Parameters for our FP survey are the following four initial cluster
conditions: $M$, $R$, the inclination of the orbital plane $i$ relative
to the Galactic plane, and the orbit eccentricity $e$, which is defined as
$(R_a-R_p)/(R_a+R_p)$ with $R_a$ and $R_p$ being the apocentre and
pericentre distances, respectively.
An initial $R$ is defined to be $R_a$ of the orbit (i.e., clusters are
initially located at $R_a$).

We choose eight $M$ values from $10^{3.5}$ to $10^7 \msun$ and
six $R$ values from $10^3$ to $10^{4.67}$~pc, both equally spaced
on the logarithmic scale.  For the orbit inclination and eccentricity,
we choose $e=0$, 0.125, 0.25, 0.5, and 0.75, and $i=15^\circ$, $45^\circ$,
and $75^\circ$, respectively.  We perform FP calculations for all
possible combinations out of these four parameters, thus the total
number of cluster models considered in our study amounts to 720.

The stellar density and velocity dispersion distributions within each cluster
follow the King model (King 1966) with a concentration parameter
$W_0=7$ and with no initial velocity anisotropy and no mass segregation.
Clusters are assumed to initially fill the tidal radius, but there is a
question regarding what tidal radius one needs to adopt for clusters with
eccentric orbits---if the star formation in a cluster takes place on a
time-scale much longer than the orbital timescale of the cluster, the tidal
radius at the pericentre would be appropriate, and if much shorter, any
value between the pericentre and the apocentre would be possible.  Here,
we assume that the clusters initially fill a tidal radius of $r_t(R_p)$
following Baumgardt (1998), but also discuss the case where the clusters
initially fill a tidal radius of $r_t(R_a)$.  For the initial stellar mass
function within each cluster, we adopt the model by Kroupa (2001) with
a mass range of 0.08--$15 \msun$, which is realized by 15 discrete
mass components in our FP model.

%%%%%%%%%%%%%%%%%%%%%%%%%%%%%%%%%%%%%%%%%%%%%%%%%%%%%%%%%%%%%%%%%%%%%%%%%%%%%%%%
\section{BEST-FIT INITIAL MF and RD}
\label{sec:bestfit}

\subsection{Evolution of individual GCs}

\begin{figure}
%Fig 1
\centerline{\epsfig{file=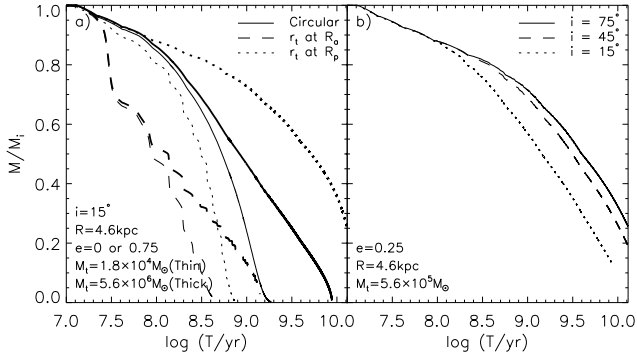,scale=0.8}}
\caption{\label{fig:cluster}Evolution of cluster mass $M$ relative to
the initial mass $M_i$ for a few clusters to show its dependence on the
orbital eccentricity ($a$) and inclination ($b$).  The values given in
the plot are initial cluster parameters.}
\end{figure}

First, we briefly discuss the effects of initial $e$ and $i$ on the evolution
of the individual clusters.  Fig. \ref{fig:cluster}(a) compares the clusters
on eccentric orbits to those on circular orbits that have a radius of $R_a$
of the eccentric orbits, for two different $M$'s.
In case the initial $r_t$ is determined at $R_a$, both massive and
light clusters on eccentric orbits evolve faster than the cluster on a
circular orbit, because the clusters on eccentric orbits are exposed to
stronger tidal fields near $R_p$.  In case the initial $r_t$ is determined
at $R_a$, both massive and light clusters on eccentric orbits also
evolve faster than the eccentric case where the initial $r_t$ is
determined at $R_p$, because for a given $M$, the cluster with an initially
larger $r_t$ would be more vulnerable to the strong tidal field near $R_p$.
However, in case
the initial $r_t$ is determined at $R_p$, the light cluster evolves
faster than the circular case while the massive cluster evolves slower
than the circular case.  This is because the light cluster has a
shorter relaxation time and thus its stars fill $r_t$ more quickly
than the massive cluster when passing near $R_a$.  When the cluster
passes near $R_p$, the light cluster then loses stars more quickly
over its shrunk $r_t$.

Fig. \ref{fig:cluster}(b) shows that clusters with $i \gsim 45 ^\circ$
evolve on similar timescales, but a cluster with a small $i$ evolves
somewhat faster due to a longer time spent while crossing the disc,
which results in a stronger disc shock.

\subsection{Synthesis of FP calculations}

As discussed in \S \ref{sec:model}, we perform a total of 720 FP
calculations with different initial cluster conditions in 4-dimensional
parameter space, $M$, $R$, $e$, and $i$.  The goal of our
present study is to find the best-fitting initial distributions of these
variables, and for this, we adopt a lognormal function
${\rm d}N(M) \propto \exp\{-0.5[\log(M/M_p)/\sigma_{\log M}]^2\} {\rm d}M$
and a truncated power-law function ${\rm d}N(M) \propto M^{-\alpha} {\rm d}M$
(only if $M \ge M_l$) for $M$ and a softened power-law function
${\rm d}N(R) \propto 4\pi R^2 {\rm d} R / [1+(R/R_0)^\beta]$ for $R$.
We assume that the initial MF is independent of $R$.
For the sake of simplicity, we do not parameterize the distributions
for $e$ and $i$, and adopt fixed isotropic distributions,
${\rm d} N(e) = 2e \, {\rm d}e$ and ${\rm d} N(i) = \sin i \, {\rm d}i$,
respectively.

Once the calculations of 720 FP models are done, the aforementioned set
of initial MF and RD models are used to search for the best-fitting initial
GC distributions in 4-dimensional parameter space: ($M_p$, $\sigma_{\log M}$,
$\beta$, $R_0$) for the lognormal MF and ($\alpha$, $M_l$, $\beta$,
$R_0$) for the power-law MF.  We synthesize our 720 FP calculations with
appropriate weights to produce a given initial MF and RD, and find a set of
($M_p$, $\sigma_{\log M}$, $\beta$, $R_0$) or ($\alpha$, $M_l$, $\beta$,
$R_0$) that best fits the present-day MF and RD of the Galactic GC system.
When finding the best-fitting initial MF and RD, we minimize the sum of
the $\chi^2$ values from both $M$ and $R$ histograms, which are
constructed by using 11 bins between $10^{3.75} \msun$ and $10^{6.5} \msun$
for $M$ and 11 bins between $10^{2.6}$~pc and $10^{4.8}$~pc for $R$,
both equally spaced in logarithmic scale.  Since the model $M$ and $R$
histograms are both constrained by the observed number of clusters in our
binning ranges, the degree of freedom for our $\chi^2$ test is
20.\footnote{We have tried a 2D Kolmogorov-Smirnov test and a $\chi^2$ test
with 2D bins as these tests can consider any correlations between $M$ and $R$,
but found that these 2D goodness-of-fit tests with a relatively small
number of observed incidences, 95, result in a rather large acceptable
ranges of parameter space.  Thus we first find the best-fitting parameters
without considering the correlation between $M$ and $R$ instead, and then
check if the best-fitting 13~Gyr model MFs have the $R$ dependence consistent
with that of the observed MF.  This way, we were able to find the model that
reproduces the observed MF and RD very well simultaneously.}

Unlike $M$, $R$ values evolve oscillating between $R_p$ and $R_a$,
and thus the model RD at 13~Gyr constructed from our 720 FP calculations
may suffer a significant random noise.  To decrease this noise, we build
the model RD by adding up the probability distribution between $R_p$ and
$R_a$ that are given by the orbital information at 13~Gyr.

\subsection{Best-fitting initial GC distributions}

\begin{figure}
%Fig 2
\centerline{\epsfig{file=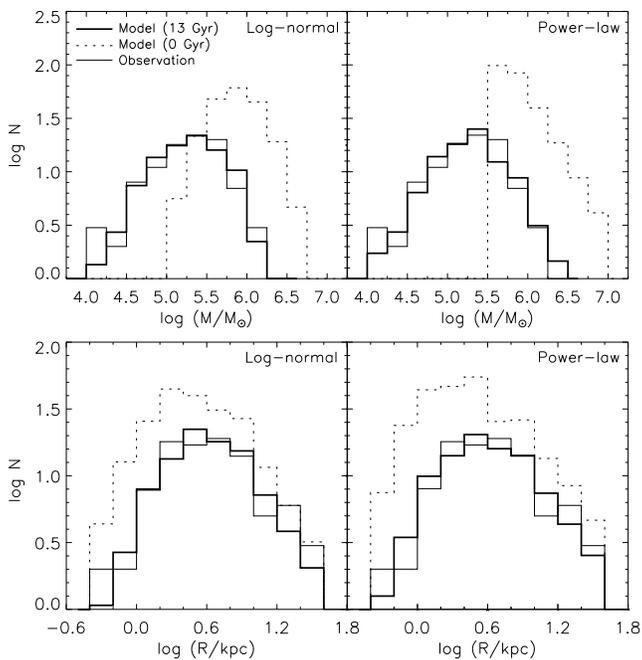,scale=0.8}}
\caption{\label{fig:bestfit}$M$ (upper panels) and $R$ (lower panels)
histograms of our best-fitting models at 13~Gyr (thick solid lines) with
a lognormal initial MF (left-hand panels) and with a power-law initial MF
(right-hand panels) along with the histograms for observed native GCs (thin
solid lines).  Also shown are the initial $M$ and initial phased-mixed $R$
histograms of our best-fitting models (dotted lines).}
\end{figure}

\begin{figure}
%Fig 3
\centerline{\epsfig{file=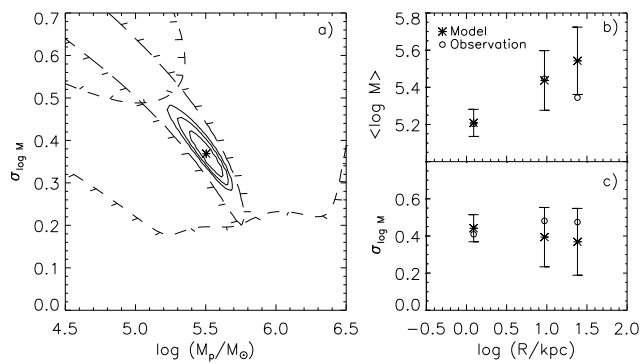,scale=0.8}}
\caption{\label{fig:chisq}Contours of $\Delta \chi^2$ values at 1, 4,
and 9 (corresponding to 1, 2, and 3$\sigma$ confidence intervals;
solid lines) from the $\chi^2$ test between our lognormal model
and the observation in the $\log M_p$--$\sigma_{\log M}$ plane (panel $a$).
Also plotted are the 5 per cent significance levels for $\chi^2(\langle \log M
\rangle)$ (long dash) and $\chi^2(\sigma_{\log M})$ (short dash), where
the perpendicular tick marks point in the downhill direction.  The minimum
$\chi^2$ point (or the best-fitting model), marked with an asterisk, is
located inside the 5 per cent significance levels of these two quantities,
implying that our best-fitting model is consistent with the $R$ dependence of
the observed MF.  $\langle \log M \rangle$ (panel $b$) and $\sigma_{\log M}$
values (panel $c$) of the best-fitting lognormal model (asterisks) and
the observation (open circles) for three radial bins.  Errorbars indicate
the 1$\sigma$ uncertainties for the model.}
\end{figure}

The best-fitting initial GC MFs and RDs from the $\chi^2$ test between our
models and the observed Galactic native (OH+BD) GCs are presented in
Table \ref{table:bestfit}.  The best-fitting models for our standard initial
condition have acceptably high $p$ values (significances), and show a good
agreement at 13~Gyr with the observed MF and RD (see Fig. \ref{fig:bestfit}).
Fig. \ref{fig:chisq}(a) shows that the confidence intervals for 1, 2 and
3$\sigma$ are formed in a relatively small region, implying that our test
gives rather small uncertainties.

\begin{table*}
%Table 1
\begin{minipage}{15cm}
\caption{\label{table:bestfit}Best-fitting parameters for initial GC distributions}
\begin{tabular}{@{}lcccccccccc}
\hline
Model & MF & $\log M_p$ & $\sigma_{\log M}$ & $\alpha$ & $\log M_l$ & $\beta$
      & $R_0$ & $M_{T,i}$ & $\chi^2$ & $p$-value \\
\hline
Standard           & L & $5.61^{+.11}_{-.15}$ & $0.33^{+.06}_{-.05}$ &&& $4.24^{+.21}_{-.21}$ & $2.9^{+.6}_{-.6}$ & $1.5\times 10^8$ & 12.9 & 88~\%\\
Standard           & P &&& $2.31^{+.09}_{-.09}$ & $5.59^{+.15}_{-.18}$ & $4.00^{+.22}_{-.18}$ & $2.0^{+.6}_{-.6}$ & $1.8\times 10^8$ & 14.8 & 79~\%\\
Circular           & L & $5.41^{+.11}_{-.18}$ & $0.48^{+.06}_{-.04}$ &&& $4.17^{+.17}_{-.18}$ & $2.0^{+.2}_{-.3}$ & $1.6\times 10^8$ & 27.4 & 12~\%\\
$\cos i\,{\rm d}i$ & L & $5.73^{+.10}_{-.09}$ & $0.29^{+.04}_{-.04}$ &&& $3.84^{+.21}_{-.21}$ & $2.0^{+.5}_{-.6}$ & $1.6\times 10^8$ & 16.2 & 70~\%\\
$r_T(R_a)$         & L & $6.08^{+.06}_{-.09}$ & $0.23^{+.05}_{-.05}$ &&& $4.50^{+.16}_{-.16}$ & $2.7^{+.2}_{-.2}$ & $6.9\times 10^8$ & 19.9 & 46~\%\\
\hline
\end{tabular}

\medskip
Best-fitting initial distribution parameters for standard and non-standard
initial conditions.  `L' in the MF column stands for the lognormal initial
MF model, and `P' for the power-law initial MF model.
The $p$ value is the probability
of having a $\chi^2$ that is larger than the value obtained from our
$\chi^2$ test between the model and the observation, whose degree of
freedom is 20.  The masses are in units of $M_\odot$ and the radii in
units of kpc.
\end{minipage}
\end{table*}

To see if our best-fitting models agree with the observed $R$ dependence
of the MF, we calculate $\chi^2$ for the differences of $\langle \log M
\rangle$ and $\sigma_{\log M}$ between the model and the observation:
\begin{eqnarray}
	\chi^2(\langle \log M \rangle)
		& = & \sum_j \frac{(\langle \log M_{o,j} \rangle -
		      \langle \log M_{m,j} \rangle)^2}
		      {\sigma_{\log M_{m,j}}^2 / N_{o,j}} \nonumber \\
	\chi^2(\sigma_{\log M}) & = & \sum_j \frac{(\sigma_{\log M_{o,j}} -
		      \sigma_{\log M_{m,j}})^2}
		      {\sigma_{\log M_{m,j}}^2 / 2N_{o,j}},
\end{eqnarray}
where subscripts $o$ and $m$ stand for the observation and the model,
respectively, and subscript $j$ represents the radial bin (we use thee radial
bins for this test).  We find that the $p$ values from the above $\chi^2$
tests for our best-fitting lognormal and power-law models are larger than 50
per cent, implying that our best-fitting models have 13~Gyr MFs whose $R$
dependence is not significantly different from the observed MF (see Fig.
\ref{fig:chisq}b and c).

Our best-fitting models have initial total masses in GCs ($M_{T,i}$) of
1.5--$1.8 \times 10^8 \msun$ and the masses that have left GCs during the
lifetime of the Galaxy ($\Delta M_T$) of 1.2--$1.5 \times 10^8 \msun$.
Our $\Delta M_T$ values are similar to or a few times larger than
previous estimates by Baumgardt (1998; 4--$9.5 \times 10^7 \msun$)
and Vesperini (1998; $5.5 \times 10^7 \msun$).  Our best-fitting models
result in the fraction of stars that remain in GCs until today of
14--18 per cent, which is comparable to the value obtained by Fall \& Zhang
(2001) for their lognormal model, 16 per cent.

In spite of the comparable $M_{T,i}$, the MF of our best-fitting lognormal
model is
located in the more massive side with a narrower width than previous studies:
it has a larger $M_p$ ($10^{5.61} \msun$) and a smaller $\sigma_{\log M}$
(0.33) than Vesperini (1998; $M_p= 10^5 \msun$, $\sigma_{\log M}=0.7$)
and Fall \& Zhang (2001; $M_p= 2 \times 10^5 \msun$, $\sigma_{\log M}=0.5$).
More various and realistic disruption mechanisms considered in our FP
models increased the $\dot M$ of individual GCs and resulted
in a larger initial $M_p$ to fit the present-day GCMF.  Since our
best-fitting model needs to fit the present-day GCMF with an increased
$M_p$, a smaller $\sigma_{\log M}$ is necessary as a compensation.

Our best-fitting lognormal model has an initial RD of $\beta=4.24$,
$R_0=2.9$~kpc,
but recall that this is the distribution of initial $R_a$, and the true
RD can be obtained by mixing the orbital phases of individual GCs.
We find that the initial phase-mixed RD (see Fig. \ref{fig:bestfit})
is well described by a softened power-law function with $\beta=3.99$,
$R_0=1.7$~kpc.
The 13~Gyr RD of our best-fitting lognormal model in the figure shows
a significant evolution from the initial value only in the small to
intermediate $R$ regions.  This shows that the GCs with smaller
$R$ values are more vulnerable to mass loss and disruption because they
have smaller relaxation times for a given $M$ and encounter disc/bulge
shocks more often.

The MF of our best-fitting power-law model has $\alpha=2.31$, $M_l=10^{5.59}
\msun$.
This $\alpha$ value is in a good agreement with the observed range of
$\alpha$ for giant molecular clouds and their star-forming cores in the
local group of galaxies (e.g., Rosolowsky 2005), 1.5--2.5.  Our best-fitting
$M_l$ is larger than that found by Parmentier \& Gilmore (2005) for their
truncated power-law IGCMF model, $\sim 10^5 \msun$.  We attribute this
difference also to the fact that our models consider more various and
realistic disruption mechanisms.  When $M_l \lsim 10^{4.5} \msun$,
our 13 Gyr MF has either too small $M_p$ or too large $\sigma_{\log M}$,
compared to the observed MF.  Our best-fitting power-law model is found to
have an initial phase-mixed RD with $\beta=3.91$, $R_0=1.3$~kpc, which
is not too different from those of the lognormal model.

\subsection{Model dependences}

\begin{figure}
%Fig 4
\centerline{\epsfig{file=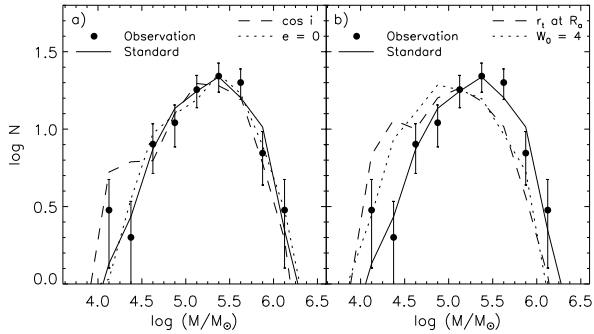,scale=0.75}}
\caption{\label{fig:vary}GCMFs at 13~Gyr for several different initial
conditions (lines)
along with the observed GCMF (filled circles).  Solid lines are our
best-fitting lognormal GCMFs with our standard initial conditions.
Each of the dashed and dotted lines shows the GCMF obtained with
one different initial condition from the standard model as labeled.}
\end{figure}

Now we discuss how the assumptions adopted in our models affect our results.
In this subsection, we concentrate on our lognormal IGCMF models only.
Fig. \ref{fig:vary}$a$ shows that the 13~Gyr GCMF does not change
significantly even when we have somewhat different initial distributions
for $e$ and $i$ (circular orbits only instead of $2e \, {\rm d}e$
and $\cos i \, {\rm d}i$ instead of $\sin i \, {\rm d}i$).
The weak dependence of the GCMF on $i$ distribution is easily expected
from the evolution of individual clusters (Fig. \ref{fig:cluster}$b$), but
the relatively weak dependence of the GCMF on $e$ distribution is a bit
surprising, considering the strong dependence on $e$ seen in the $M$
evolution of individual clusters (Fig. \ref{fig:cluster}$a$).  The latter
is thought to be because massive and light clusters have an opposite
dependence of $\dot M$ on $e$ (decreasing $\dot M$ with increasing $e$ for
massive clusters, but increasing $\dot M$ with increasing $e$ for
light clusters) and these opposite effects nearly cancel out each other.

On the other hand, the evolution of the GCMF is quite sensitive to the
choice of initial $r_t$ and $W_0$ (see Fig. \ref{fig:vary}$b$).
As seen in Fig. \ref{fig:cluster}(a), clusters evolve considerably faster
when the inital $r_t$ is determined at $R_a$ than at $R_p$, thus the
13~Gyr GCMF becomes significantly smaller in height in case of the former.
The GCMF significantly shifts toward the lower $M$ side also when the initial
$W_0$ is set to four instead of seven (when the cluster is initially less
concentrated).  This is because a cluster with a smaller $W_0$ has
relatively more stars in the outskirt compared to the core, and these
stars have more chances of escaping the cluster than those in the core.

Table \ref{table:bestfit} shows the best-fitting model parameters for our
non-standard initial conditions as well.  $M_{T,i}$ is not sensitive
to the initial $e$ or $i$ distribution, but is more than four times larger
when the initial $r_t$ is defined at $R_a$.  The mass that has left from
GCs for the latter case, $6.6 \times 10^8 \msun$, is a significant
fraction of the current mass in the stellar halo between 4 and 25~kpc,
$9 \times 10^8 \msun$ (Suntzeff, Kinman, \& Kraft 1990).  This implies that
the initial size of the GCs, along with the remnant gas expulsion in
the early stage of the GC formation (Parmentier \& Gilmore 2007), is an
important factor in determining the origin of the halo stars.

We were not able to find an initial MF and RD model that matches the
observed Galactic GC system with a $p$ value larger than 1 per cent, when the
initial $W_0$ is 4, instead of 7.  This indicates that, within the limit
of our parameterization for the initial MF and RD, a significant
portion of the GCs are likely to have formed with $W_0 \gsim 7$.

\section{SUMMARY}

We have calculated the MF and RD evolution of the Galactic GC system
using the most advanced and realistic FP model.
By simultaneously fitting both MF and RD of the observed GC system,
we found that our best-fitting models have a higher $M_p$ for a lognormal
initial MF and a higher $M_l$ for a power-law initial MF than previous
estimates, but our $M_{T,i}$'s are comparable to the previous results.
Our best-fitting power-law MF model has a $\alpha=2.2$, which is in a
good agreement with the observed range of $\alpha$ for giant molecular
clouds and their star-forming cores.
Our best-fitting lognormal and power-law models have initial phase-mixed RDs
with $\beta \simeq 4$.  Our best-fitting models, which are based on the
assumptions of isotropic $e$ and $i$ distributions, agree well not only with 
the observed MF and RD, but also with the observed $R$ dependence of the MF.
Our results are insensitive to the initial distributions of $e$ and $i$,
but are rather sensitive to how the initial
tidal radius is defined and to the initial concentration of the clusters.
If the clusters are assumed to be formed at the apocentre while filling 
the tidal radius there, $\sim 75$ per cent of the mass in the current stellar
halo could be attributed to the GCs as their origin.  This implies that
almost all of the mass in the stellar halo could be attributed to
the GCs on eccentric orbits whose initial $r_t$ is much larger than
$r_t(R_p)$, as well as to the GCs disrupted in the early stage of the
cluster formation by the remnant gas expulsion.
Finally, it appears that the clusters need to have a moderate-to-high
initial concentration ($W_0 \gsim 7$) to explain the present-day MF and RD.

\section*{Acknowledgments}

We thank Holger Baumgardt, Hansung Gim, Pavel Kroupa, and Hyung Mok Lee for
helpful discussion.  We appreciate the anonymous reviewers for their comments,
which improved our manuscript.  This work was supported by Korea Research
Foundation Grant funded by Korea Government (MOEHRD, Basic Reasearch
Promotion Fund; KRF-2005-015-C00186).  This work was, in part, supported
by the BK21 program as well.

%%%%%%%%%%%%%%%%%%%%%%%%%%%%%%%%%%%%%%%%%%%%%%%%%%%%%%%%%%%%%%%%%%%%%%%%%%%%%%%%

%\bsp
\label{lastpage}
\end{document}